\documentclass[12pt]{article}
\usepackage{amsmath,amsthm,amssymb}

\newcommand\beq{\begin{equation}}
\newcommand\eeq{\end{equation}}
\newcommand\bea{\begin{eqnarray}}
\newcommand\eea {\end{eqnarray}}

\def\hil{{\mathcal{H}}}
\def\un\a{{\underline\alpha}}
\def\la{{\langle}}
\def\ra{{\rangle}}
\def\a{{\alpha}}
\def\b{{\beta}}

\def\d{{\delta}}

\def\Tr{{\mathrm{Tr}}}

\title{Quantum Zeno Effect in the Decoherent Histories}

\author{Petros Wallden\footnote{Raman Research Institute, Theoretical Physics Group; Sadashivanagar, Bangalore - 560
080, India; on leave from: Imperial College, Theoretical Physics
Group; Blackett Laboratory, London SW7 2BZ, UK; email:
petros.wallden@gmail.com}}

\date{}

\begin{document}

\maketitle

\begin{abstract}
The quantum Zeno effect arises due to frequent observation. That
implies the existence of some experimenter and its interaction with
the system. In this contribution, we examine what happens for a
closed system if one considers a quantum Zeno type of question,
namely \emph{what is the probability of a system, remaining always
in a particular subspace}. This has implications to the arrival time
problem that is also discussed. We employ the decoherent histories
approach to quantum theory, as this is the better developed
formulation of closed system quantum mechanics, and in particular,
dealing with questions that involve time in a non-trivial way. We
get a very restrictive decoherence condition, that implies that even
if we do introduce an environment, there will be very few cases that
we can assign probabilities to these histories, but in those cases,
the quantum Zeno effect is still present.

\end{abstract}

\section{Motivation}
A remarkable property of quantum mechanics, is the so called quantum
Zeno effect \cite{Misra}. This effect, is that frequent observation
slow down the evolution of the state, with the limit of continuous
observation leading to ``freezing'' of the state\footnote{To be more
precise, restriction to a subspace.}. This has been experimentally
verified. The intuitive explanation, is that the interaction of the
observer with the system leads to this apparent paradox. It would
therefore be interesting to see whether this effect persists if we
consider a closed system. We would try to see what is the
probability of a closed system remaining in a particular subspace of
its Hilbert space with no external observer. This directly relates
to the arrival time problem as well (e.g. \cite{HaZa,Wallden2006}).
Having said that, we should emphasize that in closed systems, we
cannot in general assign probabilities to histories, unless they
decohere and it is this property that resolves the apparent paradox
that arises.

\section{This paper}

This contribution is largely based on Ref.\cite{Wallden2006}. In
Section \ref{Intro}
 we revise the quantum Zeno effect and the decoherent histories, and
 introduce a new formula for the restricted propagator that will be
 of use further. In Section \ref{Prob} we see what probabilities we
 would get if we had decoherence, that highlights the persistence of
 the quantum Zeno effect. In Section \ref{Decoh Cond} we get the decoherence
 condition that in Section \ref{Arrival} is stressed how restrictive is by considering the arrival time problem. We
 conclude in Section \ref{Concl}.
\section{Introductory material}\label{Intro}
\subsection{Quantum Zeno effect}
 In standard Copenhagen quantum mechanics,
the measurement is represented by  projecting the state to a
subspace defined by the eigenstates that correspond to the range of
eigenvalues  of the measured physical quantity. The latter is
represented by a self-adjoint operator. The state, otherwise evolves
unitarily: $\hat U(t)=\exp (-i\hat Ht)$, where  $\hat H$ is the
Hamiltonian. It is then a mathematical fact, that frequent
measurement, of the same quantity (subspace) leads to slow down of
the evolution, i.e. decreases the probability that the state evolves
outside the subspace in question. This resembles the ancient Greek,
Zeno paradox  ($ Z\acute{\eta}\nu\omega\nu$), and thus the name.

The continuum measurement limit, leads to zero probability of
leaving the observed subspace. The state continues to evolve
(unitarily), but restricted in the subspace of observation
\cite{Facchi2}. This implies that if we project to a one-dimensional
subspace, the state stops evolving. In most literature, the question
is of a particle decaying or not, so the last comment applies. In
particular, the above phenomenon is still present for  infinite
dimensional Hilbert spaces, but provided that the restricted
Hamiltonian ($H_r=PHP$) is self-adjoint, as we will see later.

\subsection{Decoherent histories}
Decoherent histories approach to quantum theory is an alternative
formulation designed to deal with closed systems and it was
developed by Griffiths \cite{Gri84}, Omn\`es \cite{Omn88a}, and
Gell-Mann and Hartle \cite{GH90b}. There is no external observer, no
a-priori environment-system split. The main mathematical aim of this
approach, is to see when is it meaningful to assign probabilities to
a history of a closed quantum system and of course to determine this
probability.

Here we will revise the standard non-relativistic quantum mechanics
in decoherent histories formulation. To each history
($\underline\alpha$) corresponds a particular class operator
$C_{\underline\alpha}$,

\beq\label{1.6} C_{\underline\a}=P_{\a_n}e^{-iH(t_n-t_{n-1})}
P_{\a_{n-1}}\cdots e^{-iH(t_2-t_1)}P_{\a_1}\eeq Where $P_{\a_1}$ etc
are projection operators corresponding to some observable, $H$ is
the Hamiltonian, and $t_n$ is the total time interval we consider.
This class operator corresponds to the history, the system is at the
subspace spanned by $P_{\alpha_1}$ at time $t_1$ at $ P_{\alpha_2}$
at time $t_2$ and so on. The probability for this history, provided
we had some external observer making the measurement at each time
$t_k$ would be

\beq\label{3.2 candidate}
p(\underline\a)=D(\underline\a,\underline\a)=\Tr(C_{\underline\a}\rho
C^\dag_{\underline\a})\eeq where $\rho$ is the initial state. In the
case of a closed system, Eq.(\ref{3.2 candidate}) fails in general
to be probability due to interference\footnote{The additivity of
disjoint regions of the sample space is not satisfied by
Eq.(\ref{3.2 candidate})}.

There are, however, certain cases where we can assign probabilities.
This happens if for a complete set of histories, they pairwise obey

\beq\label{3.2 decoherence}
D(\underline\a,\underline\b)=\Tr(C_{\underline\a}\rho
C^\dag_{\underline\b})=0\quad\forall\quad
\underline\a\neq\underline\b\eeq In that case, the complete set of
histories is called \emph{decoherent} set of histories and we can
assign to each history of this set the probability of Eq.(\ref{3.2
candidate}). In order to achieve a set of histories that satisfy
Eq.(\ref{3.2 decoherence}) in general we need to consider coarse
grained histories, or/and very specific initial state
$\rho$\footnote{Note that the interaction of a system with an
environment that brings decoherence, in the histories vocabulary, is
just a particular type of coarse graining where we ignore the
environments degrees of freedom.}.

To sum up, in decoherent histories we need to first construct a
class operators that corresponds to the histories of
interest\footnote{Note that the same classical question can be
turned to quantum with several, possibly inequivalent ways. Due to
this property, the construction of the suitable class operator is
important for questions such as for example, the arrival time or
reparametrization invariant questions.}, and then confirm that these
histories satisfy Eq.(\ref{3.2 decoherence}). Only then we can give
an answer.

\subsection{The restricted propagator}

A mathematical object that will be needed for computing the suitable
class operators, is the restricted propagator. This is the
propagator restricted to some particular region $\Delta$ (of the
configuration space) that corresponds to a subspace of the total
Hilbert space denoted by $\hil_\Delta$. The most common (but not the
most general) is the path integral definition:

\begin{equation}
g_r(x,t\mid x_0,t_0)=\int_\Delta \mathcal{D}x \exp(iS[x(t)])=\langle
x| g_r(t,t_0)| x_0\rangle
\end{equation}
The integration is done over paths that remain in the region
$\Delta$ during the time interval $[t,t_0]$. The $S[x(t)]$ is as
usual the action. The operator form of the above is given by
\cite{Halliwell:1995jh,Halliwell:2005nv}:

\begin{equation} \label{restricted operator}
 g_r(t,t_0)=\lim_{\delta
t\rightarrow 0} P e^{-iH(t_n-t_{n-1})}P\cdots P e^{-iH(t_1-t_0)}P
\end{equation}
With $t_n=t$, $\delta t\rightarrow 0$ and $n\rightarrow \infty$
simultaneously keeping $\delta t \times n=(t-t_0)$. $H$ is the
Hamiltonian operator. $P$ is a projection operator on the restricted
region $\Delta$. We therefore have

\beq g_r(x,t\mid x_0,t_0)=\la x|g_r(t,t_0)|x_0\ra\eeq Note here that
the expression Eq.(\ref{restricted operator}) is the defining one
for cases that the restricted region is not a region of the
configuration space, but some other subspace of the total Hilbert
space $\hil$. The differential equation obeyed by the restricted
propagator is:

\begin{equation} \label{restricted differential}
(i\frac{\partial}{\partial t}-H) g_r(t,t_0)=[P,H] g_r(t,t_0)
\end{equation}
Which is almost the Schr\"{o}dinger equation, differing by the
commutator of the projection to the restricted region with the
Hamiltonian.

The most useful form, for our discussion was derived in Ref.
\cite{Wallden2006}

\begin{equation} \label{restricted zeno}
g_r(t,t_0)=P\exp\left(-i(t-t_0)P H P\right)P
\end{equation}
Note that $PH P$ is the Hamiltonian projected in the subspace
$\hil_\Delta$. To prove Eq.(\ref{restricted zeno}) we multiply Eq.
(\ref{restricted differential}) with $P$ we will then get

\beq (i\frac{\partial}{\partial t}-PHP) g_r(t,t_0)=0\eeq using the
fact that $P[H,P]P=0$ and that the propagator has a projection $P$
at the final time. This is Schr\"{o}dinger equation with Hamiltonian
$PHP$. It  is evident that this leads to the full propagator in
$\hil_\Delta$ provided that the operator $PHP$ is self-adjoint in
this subspace \cite{Facchi2}\footnote{A detailed proof from
Eq.(\ref{restricted operator}) can be found in \cite{Wallden2006}.}.

\section{Quantum Zeno histories}
In this section we will examine the question \emph{what is the
probability for a system to remain in a particular subspace, during
a time interval $\Delta t=t-t_0$}. We will see the probabilities and
decoherence conditions for the general case, and then see what this
implies for the arrival time problem, which is just a particular
example.
\subsection{The class operator and probabilities}\label{Prob}

There are several ways of turning the above classical proposition to
a quantum mechanical one. The most straight forward is the
following. We consider a system being in one subspace by projecting
to that, and the history of always remaining in that subspace
corresponds to the limit of projecting to the region evolving
unitarily but for infinitesimal time and then projecting again, i.e.
taking the $\d t$ between the propositions going to zero. The class
operator for remaining always in that subspace follows from
Eq.(\ref{1.6}) by taking each $P_{\a k}$ being the same ($P$) and
taking the limit of $(t_k-t_{k-1})$ going to zero for each $k$. We
then have

\beq C_\a (t,t_0)=g_r(t,t_0)\eeq and the class operator for not
remaining at this subspace during all the interval is naturally

\beq C_\b (t,t_0)=g(t,t_0)-g_r(t,t_0)\eeq with
$g(t,t_0)=\exp(-iH(t-t_0))$ the full propagator.

Let us, for the moment, assume that the initial state $|\psi\ra$ is
such, that we do have decoherence. We will return later to see when
this is the case. The (candidate) probability is

\beq\label{4.1 zeno prob}
p(\alpha)=\la\psi|g_r^\dagger(t,t_0)g_r(t,t_0)|\psi\ra\eeq Following
Eq.(\ref{restricted zeno}) it is clear\footnote{Provided $PHP$ is
self-adjoint in the subspace. This is true for finite dimensional
Hilbert spaces and has been shown to be true for regions of the
configuration space in a Hamiltonian with at most quadratic momenta
\cite{Facchi2}.} that

\beq g_r^\dagger(t,t_0)g_r(t,t_0)=P\eeq which then implies

\beq p(\a)=\la\psi|P|\psi\ra\eeq For an initial state that is in the
subspace defined by $P$, the probability to remain  in this subspace
is one. This is the usual account of the quantum Zeno effect. As it
is stressed in other literature, to have the quantum Zeno is crucial
that the restricted Hamiltonian $H_r=PHP$ to be self-adjoint
operator in the subspace. Note, that this only states that the
system remain in the subspace, but it does not ``freeze'' completely
and in particular follows unitary evolution in the subspace with
Hamiltonian, the restricted one $H_r$. The form of
Eq.(\ref{restricted zeno}) of the restricted propagator makes the
latter comment more transparent.

\subsection{Decoherence condition}\label{Decoh Cond}

All this is well understood for open systems with external
observers. To assign the candidate probability of Eq.(\ref{4.1 zeno
prob}) as a proper probability of a closed system, we need the
system to obey the decoherence condition, i.e.

\beq D(\a,\b)=\la\psi|C^\dag_\b C_\a|\psi\ra=0\eeq and this implies
that \beq\label{4.2 decoh cond}
\la\psi|g_r^\dagger(t,t_0)g(t,t_0)|\psi\ra=\la\psi|P|\psi\ra\eeq
which is a very restrictive condition and only very few states
satisfy this, as we will see in the arrival time example. The
condition, essentially states that the overlap of the time evolved
state ($g(t,t_0)|\psi\ra$) with the state evolved in the subspace
($g_r(t,t_0)|\psi\ra$) should be the same at the times $t_0$ and
$t$. Given that the restricted Hamiltonian leads, in general, to
different evolution, the condition refers only to very special
initial states with symmetries, or for particular time intervals
$\Delta t$.

\section{Arrival time problem}\label{Arrival}

The arrival time problem is the following: \emph{What is the
probability that the system crosses a particular region $\Delta$ of
the configuration space, at any time during the time interval
$\Delta t=(t-t_0)$.} One can attempt to answer this, by considering
what is the probability that the system remains always in the
complementary region $\bar\Delta$. So if $\mathcal{Q}$ is the total
configuration space, we have $\Delta\cup\bar\Delta=\mathcal{Q}$ and
$\Delta\cap\bar\Delta=\emptyset$. Taking this approach to the
arrival time problem, the relation with the quantum Zeno histories
is apparent, since it is just the special case, where the subspace
of projection is a region of the configuration space ($\bar\Delta$)
and the Hamiltonian is quadratic in momenta, i.e.

\bea \bar P&=&\int_{\bar\Delta}|x\ra\la x|dx\nonumber\\ \hat
H&=&\hat p^2/2m+V(\hat x)\eea
 This particular case is infinite dimensional, but as shown in Ref.
\cite{Facchi2} the restricted Hamiltonian is indeed self-adjoint and
the arguments of the previous section apply.

Before proceeding further, we should point out that one could
construct different class operators that would also correspond to
the (classical) arrival time question. For example, one could
consider having POVM's\footnote{Positive Operator Valued Measure}
instead of projections at each moment of time, or could have a
finite (but frequent) number of projections (not taking the limit
where $\d t\rightarrow 0$). These and other approaches are not
discussed here.

Let us see now, what the quantum Zeno effect implies about the
arrival time. It states that a system initially localized outside
$\Delta$ will always remain outside $\Delta$ (if it decoheres) and
therefore we can only get zero crossing probabilities. This is
definitely surprising, since for a wave packet that is initially
localized in $\bar\Delta$ and its classical trajectory crosses
region $\Delta$, we would expect to get crossing probability one.
The resolution comes due to the decoherence condition as will be
argued later.

Returning to the decoherence condition Eq.(\ref{4.2 decoh cond}) we
see that there is the overlap of the time evolved state with the
restricted time evolved state. In the arrival time case, the
restricted Hamiltonian corresponds to the Hamiltonian in the
restricted region ($\bar\Delta$) but with infinite potential walls
on the boundary (i.e. perfectly reflecting). We then get decoherence
in the following four cases.

\begin{itemize}
\item[(a)] The initial state $|\psi\ra$ is in an energy eigenstate,
and it also vanishes on the boundary of the region.

\item[(b)] The restricted propagator can be expressed by the method
of images\footnote{Note that the restricted propagator can be
expressed using the method of images, if and only if there exist a
set of energy eigenstates, vanishing on the boundary, that when
projected on the region $\bar\Delta$ forms a dense subset of the
subspace $\hil_{\bar\Delta}$, i.e. span $\hil_{\bar\Delta}$. This is
equivalent with requiring that the restricted energy spectrum (i.e.
spectrum of the restricted Hamiltonian $H_r$) is a subset of the
(unrestricted) energy spectrum, which is not in general the case.}
and the initial state shares the same symmetry.

\item[(c)] The full unitary evolution in the time interval $\Delta
t$ remains in the region $\bar\Delta$.

\item[(d)]Recurrence: Due to the period of the Hamiltonian and
the restricted Hamiltonian their overlap happens to be the same
after some time  $t$ as it was in time $t_0$. This depends
sensitively on the time interval and it is thus of less physical
significance.

\end{itemize}
It is now apparent that most initial states do not satisfy any of
those conditions. In particular, the wavepacket that classically
would cross the region $\Delta$, will not satisfy any of these
conditions, and we would not be able to assign the candidate
probability as a proper one, and thus we avoid the paradox. The
introduction of an interacting environment to our system, (that
usually produces decoherence by coarse-graining the environment)
does not change the probabilities and contrary to the intuitive
feeling, it does not provide decoherence for the particular type of
question we consider. This still leave us with no answer for any of
the cases that the system would classically cross the region. The
latter implies, that the straight forward coarse grainings we used,
were not general enough to answer fully the arrival time question
\footnote{For more details, examples and discussion see Ref.
\cite{Wallden2006}.}.

As a final note, we should point out that the quantum Zeno effect in
the decoherent histories, has implications for the decoherent
histories approach to the problem of time (e.g. Refs.
\cite{Halliwell:2005nv,Wallden2006}).

\section{Conclusions}\label{Concl}

We examined the quantum Zeno type of histories of a closed system,
using the decoherent histories approach. We show that the quantum
Zeno effect is still present, but only for the very few cases that
we have decoherence. The situation does not change with the
introduction of interacting environment. We see that while in the
open system quantum Zeno, the delay of the evolution arises as
interaction with the observer, in the closed system we have the
decoherence condition ``replacing'' the observer and resolving the
apparent paradox.

\paragraph{Acknowledgments:} The author is very grateful to Jonathan J. Halliwell for many
useful discussions and suggestions, and would like to thank the
organizers for giving the opportunity to give this talk and hosting
this very interesting and nice conference.


\begin{thebibliography}{99}
\bibitem{Misra}
B. Misra and E.C.G. Sudarshan, J. Math. Phys. {\bf 18}, 756 (1977).
\bibitem{HaZa} J.J.Halliwell and E.Zafiris, {Phys.Rev.} {\bf D57},
3351 (1998).
\bibitem{Wallden2006} P. Wallden, gr-qc/0607072.
\bibitem{Facchi2} P. Facchi, S. Pascazio, A. Scardicchio, and L. S. Schulman,
Phys. Rev. A \textbf{65}, 012108 (2002).
\bibitem{Gri84}
R.B. Griffiths,
\newblock  J. Stat. Phys., {\bf36} 219, (1984).

\bibitem{Omn88a}
R.~Omn\`es.
\newblock  J. Stat. Phys. {\bf 53}, 893 (1988).

\bibitem{GH90b}
M.~Gell-{M}ann and J.~Hartle,
\newblock in {\em Complexity, Entropy and the Physics of
Information, SFI Studies in the Science of Complexity, {Vol. VIII}},
edited by W.~Zurek, (Addison-Wesley, Reading, 1990).

\bibitem{Halliwell:1995jh}
  J.~J.~Halliwell,
  quant-ph/9506021.

  \bibitem{Halliwell:2005nv}
  J.~J.~Halliwell and P.~Wallden,
  Phys.\ Rev.\ D {\bf 73} 024011 (2006), gr-qc/0509013.

\end{thebibliography}
\end{document}